\documentclass{article}
\usepackage{spconf,amsmath,graphicx}
\usepackage{hyperref}
\usepackage{tabularx}
\usepackage{svg}

\title{A multi-stage low-latency enhancement system for hearing aids}
\name{Chengwei Ouyang, Kexin Fei, Haoshuai Zhou, Congxi Lu, Linkai Li}
\address{Orka Inc.\\
\{chengwei, feikexin, haoshuai, chauncey, linkai\}@hiorka.com}
\begin{document}
%
\maketitle
\begin{abstract}
This paper proposes an end-to-end system for the ICASSP 2023 Clarity Challenge. In this work, we introduce four major novelties: (1) a novel multi-stage system in both the magnitude and complex domains to better utilize phase information; (2) an asymmetric window pair to achieve higher frequency resolution with the 5ms latency constraint; (3) the integration of head rotation information and the mixture signals to achieve better enhancement; (4) a post-processing module that achieves higher hearing aid speech perception index (HASPI) scores with the hearing aid amplification stage provided by the baseline system.
\end{abstract}
\section{Introduction}
\label{sec:intro}

\begin{figure}[htbp]
\begin{center}
\includegraphics[width=8.5cm]{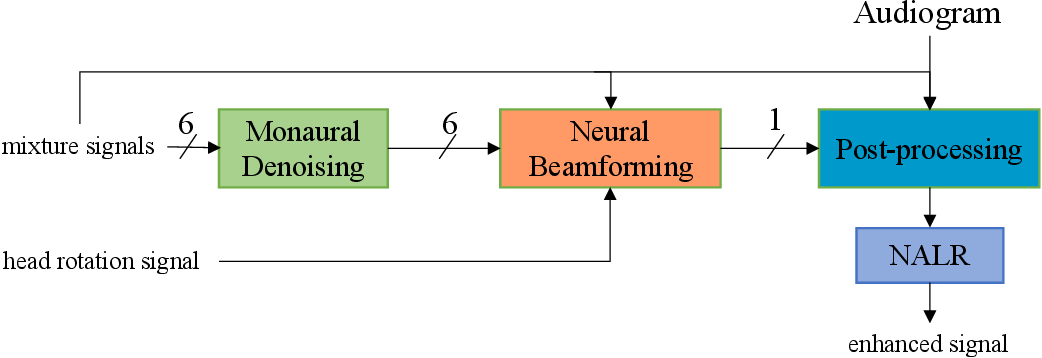}
\end{center}
\caption{Overall architecture of the end-to-end model.}
\label{fig:overall_img}
\end{figure}

The ICASSP Signal Processing Grand Challenge: Clarity Challenge (Speech Enhancement for Hearing Aids) 2023 \cite{claritychallenge} aims to improve the performance of hearing aids for speech-in-noise. This paper describes our system,  and the overall architecture is depicted in Fig. \ref{fig:overall_img}. Our system comprises three main components: a monaural denoising module and a neural beamforming module for speech enhancement, and a post-processing module for better hearing loss compensation. The system operates with a 32kHz sampling rate and utilizes a short-time Fourier transform (STFT) with a 16ms window and 2ms stride, but normally a system with this configuration has latency of 16ms. To address this issue, we propose an asymmetric window pair, which is illustrated in Section 2.1 and can effectively reduce the algorithmic latency to 4ms.

\section{Method}
\label{sec:format}

\subsection{Asymmetric Window Pair}
\label{ssec:subhead1}
The proposed asymmetric window pair is composed of a forward window $w_1[n]$ and a backward window $w_2[n]$, which is defined as
$$
w_1[n]=\begin{cases}
sin^2(\frac{n\pi}{2N_1}),\hphantom{-\text{xxxxxxx}}if 0\leq n<N_1 \\
1,\hphantom{-\text{xxxxxxxxxxxx}} \hphantom{-\text{x}}if N_1\leq n\leq N_2 \\
sin(\frac{\pi(N_2+R - n)}{4R}), \hphantom{-\text{x,}}if N_2<n\leq N_2+R
\end{cases}$$
$$w_2[n]=\begin{cases}
0,\hphantom{-\text{xxxxxxxxxxxxxx,,}} if 0\leq n<N_2-R \\
cos^2(\frac{\pi(n-N_2)}{2R}),\hphantom{-\text{xxx,}} if N_2-R\leq n\leq N_2 \\
sin(\frac{\pi(N_2 + R - n)}{2R}),\hphantom{-\text{,,}} if N_2<n\leq N_2+R
\end{cases}$$
where $R>0$ is the hop size, and $0<N_1<N_2-R$.

In most real-time speech processing system, the reconstruction latency is equal to window size, so the window size has to be short in order to achieve a low latency. But a small window size restrains frequency resolution and thus the model performance. To solve this problem, we designed a forward and backward window pair to simultaneously achieve a high frequency resolution while still maintaining a low reconstruction latency. The overall latency is independent of window size and can be two times of hop size. Besides, this window pair satisfies the Constant Overlap-Add (COLA) property and thus can achieve perfect reconstruction of the original signal. In our experiment, we set $N_1=64$, $N_2=448$ and $R=64$ under 32kHz sampling rate, which corresponds to 2ms hop size with 16ms window size. The overall latency of 4ms meets the 5ms latency requirement.

\subsection{System Description}
\label{ssec:subhead3} 
The challenge provides a dataset consisting of 6000 scenes for training, and 2500 and 3000 scenes for validation and evaluation respectively. Each scene comprises a six-channel behind-the-ear (BTE) device recordings and a head rotation signal. However, We observed that the anechoic target signals were slightly misaligned with the mixture signals, which caused errors when using  networks and loss function in either the time or the complex domain. As the challenge did not provide any means to distinguish between targets and interferers for a monaural system, so it relied heavily on multi-channel and phase information. We found that combining a monaural network operating in magnitude domain with a multi-channel network operating in complex domain yielded superior performance. For the monaural denoising network, we employed DNN-Net from \cite{SDD-NET} as our magnitude domain network. We separated the input mixture signals into magnitude and phase components and rearranged the channel dimension into batches to minimize any potential interference from the phase components. For the neural beamforming network, we utilized SAF module from \cite{s-dccrn} as our complex domain network and extended it to a multi-channel version with seven input channels (six channels for mixture signals and one channel for head rotation information).

In this challenge, the provided hearing aid system employs the National Acoustic Laboratories (NALR) fitting algorithm \cite{nalr} to achieve listener-specific amplification. However, this approach may not be effective for individuals with severe hearing loss and can result in suboptimal performance. To address this issue, we utilize a DNN configured identically to our monaural denoising module as a post-processing module. This module aimed to fine-tune the enhanced spectrogram based on the listener's audiogram to maximize speech intelligibility for individuals with hearing impairment. It takes inputs from the neural beamforming module and the listener’s audiogram, and a fully connected layer is used to combine them before being fed into the network.

\section{Experiments and Results}
\label{sec:pagestyle}

All signals are resampled to 32kHz for training. The training process is divided into two stages. In the first stage, both the monaural denoising module and neural beamforming module are trained using a multi-resolution loss function with 128-, 256-, 512-, 1024-, 2048-sample windows to compute L2 loss on magnitude. In the second stage, the model from the first statge is loaded as a pre-trained model and a post-processing module is added to train the final model. 
The NALR fitting algorithm is modified to make it differentiable and the loss function used in this stage combines differentiable HASPI loss with multi-resolution loss. The HASPI loss is computed after NALR amplification while the multi-resolution loss is computed prior to it. The purpose of using multi-resolution loss is to compress amplitude energy in output signals since HASPI loss tends to increase energy in output signals and the hearing aid speech quality index (HASQI) is highly sensitive to amplitude.

Two separate networks are trained to estimate the target anechoic signal for both ears. The Adam optimizer is used with a learning rate of 0.0003 and a batch size of 2. The final model has 7 million training parameters and requires approximately 86G FLOPs. The results presented in Table 1 demonstrate promising HASPI and HASQI scores, particularly in light of the challenge's fixed NALR fitting algorithm, which has been found to introduce significant deviations owing to its imperfect hearing loss compensation capabilities. Incorporating head rotation information also brings a small but noticeable improvement on HASPI score with minimal additional computational cost.

\begin{table}
    \centering
    \begin{tabular}{lcccc}
    \hline
         Approaches & Dataset & HASPI & HASQI & Average  \\
        \hline
        noisy & dev & 0.089 & 0.063 & 0.076  \\
        baseline & dev & 0.239 & 0.132 & 0.185  \\
        Neural beamforming & dev & 0.692 & 0.287 & 0.490  \\
          $+$Monaural denoising & dev & 0.744 & 0.356 & 0.550  \\
          $+$Post-processing & dev & 0.795 & 0.381 & 0.588  \\
          $+$Head rotation signal & dev & 0.812 & 0.392 & 0.602  \\
        \hline
         Submitted system & eval1 & 0.835 & 0.393 & 0.614  \\
          $+$Head rotation signal & eval1 & 0.838 & 0.393 & 0.616  \\
          \hline
          Submitted system & eval2 & 0.256 & 0.104 & 0.180  \\ 
         $+$Head rotation signal & eval2 & 0.257 & 0.103 & 0.180  \\
         \hline
    \end{tabular}
    \caption{Results on development and evaluation set}
    \label{tab:my_label}
\end{table}

\section{Conclusion}
\label{sec:majhead}

In this work, we proposed a deep learning based system for the ICASSP 2023 Clarity Challenge that comprises a monaural denoising module, a neural beamforming module and a post-processing module. A critical component of the system is the proposed asymmetric window pair which achieves both high frequency resolution and low reconstruction latency while satisfying the 5ms latency constraint. The experimental results show that our system significantly outperforms baseline in terms of both HASPI and HASQI score and ranked TOP 5 in the ICASSP 2023 Clarity Challenge.

\bibliographystyle{IEEEbib}
\bibliography{strings,refs}

\end{document}